\documentclass[fleqn,twoside]{article}
\usepackage{espcrc2}

\usepackage{graphicx}
\usepackage[figuresright]{rotating}

\newcommand{\AmS}{{\protect\the\textfont2
  A\kern-.1667em\lower.5ex\hbox{M}\kern-.125emS}}

\title{NuInt05 Session 2 Summary: New Experimental Results}

\author{L.~Ludovici\address{University of Rome La Sapienza and INFN,
        I-000185 Rome, Italy}, 
        K.~S.~McFarland\address{University of Rochester, Rochester, NY
        14627 USA}, 
        M.~Shiozawa\address{Kamioka Observatory, Institute for Cosmic
        Ray Research, University of Tokyo, Kamioka, Gifu 506-1205, Japan},
        G.~P.~Zeller\address{Columbia University, Department of Physics,
        New York, NY 10027 USA}}
       
\begin{document}
\begin{abstract}

We summarize the discussion and presentation of new experimental results in 
low energy neutrino scattering at the NuInt05 workshop. In addition, we 
comment on several investigations that we would like to see targeted in 
preparation for the next NuInt workshop.
\vspace{1pc}
\end{abstract}

\maketitle

\section{New Experimental Results}

We were fortunate that all of the presentations in the NuInt05 second
session showcased new or updated experimental results from a variety 
of experiments. In the case of charged current (CC) single $\pi^+$ production, this included 
the first experimental measurement of CC coherent pion production at low 
energy in the K2K SciBar detector~\cite{tanaka} and the first measurement of 
the ratio of CC $1\pi^+$ to CC quasi-elastic (QE) cross sections on a nuclear target at low energy from 
MiniBooNE~\cite{wascko}. NOMAD presented their new measurements of both 
QE and inclusive CC cross sections on carbon~\cite{petti}. 
In addition, MINOS just started taking beam data a few months before the 
workshop and presented their first near detector neutrino event 
distributions~\cite{belias}. In the neutral current (NC) sector, K2K 
reported the first observation of neutrino-induced photons from nuclear 
de-excitation with the goal of measuring a NC elastic/QE cross 
section ratio on water~\cite{kameda}. The session ended with 
presentation of the first QE neutrino data in liquid argon from
exposure of the small ICARUS prototype detector in the CERN WANF 
beam~\cite{curioni}.\\

In many cases, these presentations added data in new energy regions or
on new target materials. While we are clearly benefitting from this new data, 
these results certainly raise new questions. Many of these questions can be 
answered with data that is currently in hand or that will be collected in the 
near future. To ensure further progress and resolution of these questions, 
we present the issues and questions to address in the next NuInt workshop. 

\section{Low $Q^2$ ($Q^2 < 0.2$ GeV$^2$)}

Since the beginning of the NuInt workshop series, we have seen significant
and persistant disagreement at low $Q^2$ between model predictions and data 
from multiple experiments. This is perhaps not very surprising as this is 
the region where nuclear corrections are largest and most challenging to 
model. \\

Throughout the course of the NuInt workshops, we have learned that both the 
K2K and MiniBooNE data samples exhibit very similar $Q^2$ behavior; also,
a larger low $Q^2$ deficit is observed in single pion than in QE data. We 
propose a detailed, quantitative comparison between the low $Q^2$ behavior 
in all of available CC exclusive samples on various targets and from multiple 
experiments. This includes comparisons between QE, non-QE, and exclusive 
CC $1\pi^+$, CC $1\pi^0$, and CC multi-$\pi$ data. It would also be interesting 
to compare the K2K and MiniBooNE $^{12}$C and $^{16}$O data to data on heavier 
targets such as $^{56}$Fe (MINOS) and $^{207}$Pb (CHORUS). \\

At this workshop, K2K presented new results~\cite{tanaka} indicating that 
CC coherent $\pi^+$ production cross secton calculations are much too large, 
hence explaining the observed deficit of data events at low $Q^2$. 
This analysis selects a coherent-rich sample by requiring events have
low $Q^2$, $<0.1$~GeV$^2$, and by removing events consistent with QE
kinematics or which have extra activity at the vertex.
While 
evidence from the K2K analyses is compelling, it is important to definitively 
settle the low $Q^2$ issue with complementary data. 
This includes supporting evidence from other kinematic distributions 
(e.g., $\cos\theta_\pi$, and $t$) as well as cross-checks against CC $1\pi^0$ 
data where there is no expected coherent scattering contribution. In addition,
we propose an explicit search for CC coherent pion production in
existing data from MiniBooNE, 
to be compared to the K2K results. \\

Another interesting avenue of investigation is to check,
with present data, whether we can claim consistency between observed levels of 
coherent pion production in both of the CC $\pi^+$ and NC $\pi^0$ data samples 
or whether we are seeing differences between NC and CC coherent pion production.
Finally, antineutrino coherent pion production is also of great interest because coherent 
pion production cross sections are predicted to be similar for neutrino and 
antineutrino scattering, and therefore such a deficit should be even more pronounced in 
antineutrino single pion data where resonance pion production backgrounds
are smaller.

\section{QE Scattering}

While we are enjoying the arrival of new, higher statistics QE data from K2K, 
MiniBooNE, and NOMAD, is it really clear that we better understand QE 
scattering on nuclear targets as a result? At this workshop, it was pointed out that
K2K and NOMAD are extracting significantly different axial-vector mass ($M_A$)
parameters from a dipole fit to the QE axial form factor. To assess how meaningful these
differences are, it is imperative that we establish a common formalism with
which to directly compare experimental results. With identical nuclear models
and modern vector form factors~\cite{bradford}, we need to reassess the range 
of $M_A$ values extracted from modern QE data on a variety of targets 
($^{12}$C, $^{16}$O, $^{56}$Fe, $^{40}$Ar). What impact do different levels of 
backgrounds, efficiencies, and selection biases have on these results? 
How do the results obtained from rate versus shape ($d\sigma/dQ^2$) fits 
compare? What do similar fits to single pion distributions yield for $M_A$? \\

Can more information be added? Is there anything to be gained by refitting
old neutrino data on heavy targets? What are the prospects for 
MiniBooNE antineutrino QE data and $M_A$ fits? Can similar fits be performed
on $^{56}$Fe at MINOS~\cite{belias} or with liquid argon 
data~\cite{curioni}? The current state of affairs demands a concerted effort 
to reassess QE (and more specifically, axial-vector) parameters measured
from modern heavy target data.

\section{New Data}

We also encourage the collection of new data sets.
It is also important that we continue to build up our body of knowledge
on low energy neutrino cross section by adding new experimental data.
This new data may, of course, help clarify some of the questions being 
raised by the current datasets or have the potential to raise new ones.
Specifically, CC $1\pi^0$ data can aid in disentangling resonant and 
coherent pion production at low $Q^2$.  Studies of coherent scattering
at MINERvA~\cite{mcfarland} and theoretical studies~\cite{paschos}
indicate that kinematic separation of resonant scattering from
coherent scattering may be easier at higher energies.  Samples of $\nu_e$ QE interactions 
serve as a direct check of the predicted $d\sigma/dQ^2$ distributions 
in $\nu_e$ versus $\nu_\mu$ events. Also, any new information on antineutrino 
scattering and scattering from other nuclear targets at both higher
and lower $A$ than carbon or oxygen are, of course, always welcome. 

\section{New Theoretical Models}

It is essential that we maintain a continued effort to pit modern neutrino 
data against modern theoretical models. Theories advance most quickly
when confronted with data on an equal footing. At this workshop,
we have seen many examples of new theoretical calculations that could help
experimentalists move beyond their antiquated Fermi Gas and Rein-Sehgal based
Monte Carlos. At the same time, experiments need actual code as well as 
guidance on how to implement new theoretical predictions in regions where 
they are valid. To move forward, we encourage continued communication between
theorists and experimentalists toward this end.

\section{Conclusions}

We challenge the next NuInt workshop to explicitly target the 
questions being raised by current experimental data. Specifically:

\begin{itemize}
 \item How well do we understand the axial-vector form factor 
       contribution to QE and resonant scattering?
 \item Do we have a consistent framework to compare results from different
       experiments?
 \item Do we truly understand the behavior at low $Q^2$ in all of our
       data samples?
 \item Do we see any obstacles to incorporating new 
       theoretical calculations into our experimental descriptions?
 \item Are there any other data samples that should be studied
       in addition to those that have been reported in this and previous 
       NuInt workshops?
\end{itemize}

\noindent
We hope that we can continue to make progress by isolating new data samples 
and working toward detailed comparisons between various experimental results. 
It is clear that we need to forge ahead on all fronts in order to fully 
clarify these remaining issues.

\section{Acknowledgements}

The Session 2 convenors sincerely thank Makoto Sakuda and Okayama 
University for graciously hosting this workshop.  
Intellectual discourse never blossoms more fruitfully than it did in Okayama in
the inspiring presence of the perfect grapes.



\begin{thebibliography}{9}

\bibitem{tanaka} H.~Tanaka, these proceedings.
\bibitem{wascko} M.~O.~Wascko, these proceedings.
\bibitem{petti} R.~Petti, these proceedings.
\bibitem{belias} A.~Belias, these proceedings.
\bibitem{kameda} J.~Kameda, these proceedings.
\bibitem{curioni} A.~Curioni, these proceedings.
\bibitem{bradford} R.~Bradford, these proceedings.
\bibitem{mcfarland} K.~McFarland, these proceedings.
\bibitem{paschos} E.A.~Paschos, these proceedings.

\end{thebibliography}
\end{document}